\documentclass[%
 reprint,
 amsmath,amssymb,
 aps,
pre,
]{revtex4-1}

\usepackage{graphicx}
\usepackage{dcolumn}
\usepackage{bm}

\begin{document}

\preprint{APS/123-QED}

\title{Culinary evolution models for Indian cuisines}

\author{Anupam Jain}
 \affiliation{%
 Center for System Science, Indian Institute of Technology Jodhpur, Jodhpur, 342011, India
}%
\author{Ganesh Bagler}%
\email{bagler@iitj.ac.in}
 \affiliation{%
 Center for Biologically Inspired System Science, Indian Institute of Technology Jodhpur, Jodhpur, 342011, India
}%

\date{\today}

\begin{abstract}
Culinary systems, the practice of preparing a refined combination of ingredients that is palatable as well as socially acceptable, are examples of complex dynamical systems. They evolve over time and are affected by a large number of factors. Modeling the dynamic nature of evolution of regional cuisines may provide us a quantitative basis and exhibit underlying processes that have driven them into the present day status. This is especially important given that the potential culinary space is practically infinite because of possible number of ingredient combinations as recipes. Such studies also provide a means to compare and contrast cuisines and to unearth their therapeutic value. Herein we provide  rigorous analysis of modeling eight diverse Indian regional cuisines, while also highlighting their uniqueness, and a comparison among those models at the level of flavor compounds which  opens up molecular level studies associating them especially with non-communicable diseases such as diabetes.

\begin{description}
\item[PACS numbers]
89.75.-k, 82.20.Wt, 87.18.Vf, 87.10.Vg, 89.90.+n
\end{description}
\end{abstract}

\pacs{Valid PACS appear here}
\maketitle


\section{\label{sec:level1}Introduction \\}

Culinary systems are examples of complex dynamical systems. Culinary practices and hence food preparation procedures (recipes) have evolved to the present day traditional cuisines by tuning them so as to suit human sensibilities. Knowing the complexity of culinary evolution the question is whether it could be modeled to identify key elements that drive its nature.

Lately, culinary science has attracted the attention of physicists due to invariant patterns observed across cuisines as well as owing to cuisine-specific features that highlight evolutionary mechanisms whose understanding facilitate various applications~\cite{Kinouchi2008a,Ahn2011,Jain2015}. Understanding the process of culinary evolution can help bring to the surface, the guiding principles behind the development of the cuisine.

While the choice of food ingredients and their combinations is dictated by a range of factors such as geography, culture, climate and genetics~\cite{Pollan2014,Appadurai2009,Sherman1999,Zhu2013,Birch1999,Ventura2013a}, the sensory mechanisms of taste (gustatory) and smell (olfactory) play a dominant role to lock-in ingredients into recipes~\cite{Breslin2013}. 

Various cuisines have been reported to have generic nature of recipe size distribution and frequency-rank distribution~\cite{Zhu2013,Varshney2013}. Yet these cuisines are unique in preferring certain ingredients and ingredient combinations.
India has had a long culinary history and is characterized by diverse geographies and climates. While the cuisine of India could be seen as a single cuisine, it is represented by a blend of regional cuisines. 

Earlier studies have reported most cuisines to have positive food pairing i.e. they tend to use ingredient pairs of similar flavor and taste~\citep{Ahn2011}, Indian cuisine in contrast is reported to have negative food pairing i.e. Indian recipes tend to have ingredients of complementary flavors~\cite{Jain2015}. This is an important distinction which, apart from reflecting geoclimatic and cultural differences, says a lot about the trajectory that Indian cuisine has followed and perhaps bears specific culinary milestones in specifying recipe compositions.

Herein we ask the following questions in the context of Indian regional cuisines. 
(a)	Could a model generate a cuisine which is similar to the real world not only in terms of its statistical patterns but also in terms of its flavor profile?
(b)	What are the dominant factors in shaping the recipe size distribution?

Towards addressing the above questions we created models of culinary evolution:
(a)	Copy-mutate model – fitness random (CM-fitness random)
(b)	Copy-mutate model – fitness ranked (CM-fitness ranked)
(c)	Copy-mutate-add-delete model (CMAD model).
CM-random serves as a null model and when compared with CM-ranked, indicates the role of ingredient rank (frequency of use) in food pairing pattern.

In the second section we describe the data acquisition and correction methods with corresponding final statistics. The third section explains our model's methodology and parameters involved. It also explains the modality of authenticity study, carried out to find the most legitimate ingredients belonging to each regional cuisine. Then we provide results on reproduction of certain statistical features of the Indian cuisine (and its regional cuisines). 

\section{Data acquisition and statistics}
We began with extracting data for Indian cuisine from the website tarladalal.com (November, 2014)~\cite{Dalal2014} which is the largest online repository of recipes for Indian cuisine. After curating the data for redundant characters and words (such as contributor's name) from recipes, we were left with 2543 recipes and their corresponding ingredients. The ingredients being listed in different spellings and usage amounts and forms (chopped, sliced etc.) were required to be aliased separately after which we had 194 ingredients for the whole cuisine. Since the intent of current study was extended beyond simple statistical modeling of cuisine to look for flavor patterns within the model cuisine, we also gathered information of flavor compounds present in the ingredients and ended up with an overall list of 1170 flavor compounds corresponding to above 194 ingredients through existing data~\citep{Ahn2011,Jain2015} and resources of flavor compounds~\cite{Burdock2010}.  Table~\ref{tab:regional_cuisines} lists statistics of recipes and ingredients in each of the regional cuisines.

\begin{table}[b]
\caption{\label{tab:regional_cuisines}%
Statistics of recipes and ingredients in regional cuisines.}
\begin{ruledtabular}
\begin{tabular}{lcdr}
\textrm{Cuisine}&
\textrm{Recipe count}&
\textrm{Ingredient count}\\
\colrule
Bengali & 156 & 102\\
Gujarati & 392 & 112 \\
Jain & 447 & 138 \\
Maharashtrian & 130 & 93 \\
Mughlai & 179 & 105 \\
Punjabi & 1013 & 152 \\
Rajasthani & 126 & 78 \\
South Indian & 474 & 114 \\
\end{tabular}
\end{ruledtabular}
\end{table}

\section{The culinary evolution models}
For the purpose of random copy-mutate model~\cite{Kinouchi2008a}, we assign a number selected uniformly randomly from the range [0, 1] to each available ingredient as its `fitness' value. The meaning of this fitness value in the real world can be taken to be a quantifying parameter describing the possible preferential efficacy of an ingredient based on factors such as availability, nutritional aspect, relative popularity, flavor and cost~\cite{Kinouchi2008a}.

The copy-mutate algorithm begins by creating a seed pool, $R_0$ of 20 recipes generated by random selection of $S=7$ ingredients for each such recipe from an initial random pool $I_0$ of 10 ingredients. Further at each time step we selected a recipe randomly from the pool as `mother' recipe and made a copy of it for mutation. Within the copied recipe we chose an ingredient (of fitness $f_i$) randomly and compared its fitness value $f_i$ with the fitness value $f_j$ of another ingredient from the ingredient pool, also chosen randomly. If $f_j > f_i$ we replace the old ingredient ($i$) with this new one ($j$). Thus the copied recipe is mutated 1 time. This process of mutation is carried out $M$ number of times after which the mutated copy recipe is added back to the pool as another possible candidate of being a mother recipe in next time step.

To introduce new ingredients we also check and maintain at each time step a ratio $r$ of size of ingredients pool and size of recipe pool. The value of $r$ for current study was taken to be this ratio, calculated from empirical data. If the ratio falls below the required threshold then new ingredients are introduced in the pool by random selection from the overall available list of ingredients. 

The overall process of recipe selection-mutation is repeated till we get $R$ number of recipes which is equal to the empirical recipe count of 2,543. For normalization purposes, we create 24 such sets of random copy-mutate recipes and study overall statistics over average of all sets.

We implemented three different models:

\begin{itemize}
\item \textbf{Copy-mutate Fitness Random}\\
In this model, the `fitness' values are assigned to ingredients on a uniform random basis. This model starts with no a priory basis or bias about the fitness of certain ingredients.

\item \textbf{Copy-mutate Fitness Ranked}\\
In this model, an ingredient is assigned `fitness' value based on its empirical frequency. Thus, an ingredient with higher frequency in real world cuisine would have a higher fitness value. Obviously this model depends on fitness of an ingredient that is ascertained retrospectively.

\item \textbf{Copy-mutate-add-delete Model}\\
Going further from previously models, where the size of the recipes in a cuisine is fixed, we generated another model that has a provision for addition and deletion of an ingredient. In this model, an additional factor was introduced to choose an ingredient for addition, deletion and mutation at each time-step. 
\end{itemize}

\section{Ingredient authenticity}
Further, In order to understand and highlight the differences among regional cuisines and uniqueness of each one, we carried out a study on finding the most authentic ingredients. This study highlights ingredients more commonly used in one cuisine as compared to other cuisines. In order to compute this, we use the prevalence~\citep{Ahn2011} $P_i^c$ of an ingredient $i$ in a cuisine $c$ as $P_i^c=n_i^c/N_c$ where $n_i^c$ is the number of recipes that contain the particular ingredient $i$ in the cuisine and $N_c$ is the number of recipes in the cuisine. The relative prevalence $p_i^c$ measuring the authenticity of the ingredient $i$ is computed as the difference between the prevalence of $i$ in cuisine $c$ and the average prevalence of $i$ in all other cuisines.

\section{Results}
All eight regional cuisines reflect the culinary diversity of Indian culture. This is not only evident by the recipes belonging to each cuisines but also by the pattern of ingredient usage. This could be observed by looking at the most authentic ingredients for regional cuisines. A list of top 5 most authentic ingredients for each of the regional cuisines is given in table~\ref{tab:authentic_ingreds}.

\begin{table*}[!]
\caption{\label{tab:authentic_ingreds}%
Top 5 most authentic ingredients for each of the regional cuisine.}
\begin{ruledtabular}
\begin{tabular}{ c | c | c | c | c | c | c | c }
\textrm{Bengali}&
\textrm{Gujarati}&
\textrm{Jain}&
\textrm{Maharashtrian}&
\textrm{Mughlai}&
\textrm{Punjabi}&
\textrm{Rajasthani}&
\textrm{South Indian}\\
\colrule
coriander & asafoetida & butter & turmeric & milk & garam masala & ghee & curry leaf \\
egg plant & green bell pepper & corn grit & coconut & cardamom & wheat & fennel & black bean \\
turmeric & sesame seed & banana & cayenne & ghee & sunflower oil & cayenne & black mustard seed oil \\
milk & black mustard seed oil & tomato & cinnamon & cream & cottage cheese & chickpea & rice \\
ginger garlic paste & chickpea & corn & clove & clove & onion & cumin & tamarind \\
\end{tabular}
\end{ruledtabular}
\end{table*}

\subsection{Frequency-rank distribution}
While the authenticity study highlights uniqueness of each regional cuisine, the generic nature of frequency-rank distribution has been shown to be a rather interesting statistical feature of cuisines around the world. Its consistent nature across regional Indian cuisines has been shown earlier~\citep{Jain2015}, making it a feature of special interest and an indicator of generic culinary evolution mechanism. We began our study by adopting the model for the purpose of reproducing this pattern.

\begin{figure}
\begin{center}
\includegraphics[scale=.1]{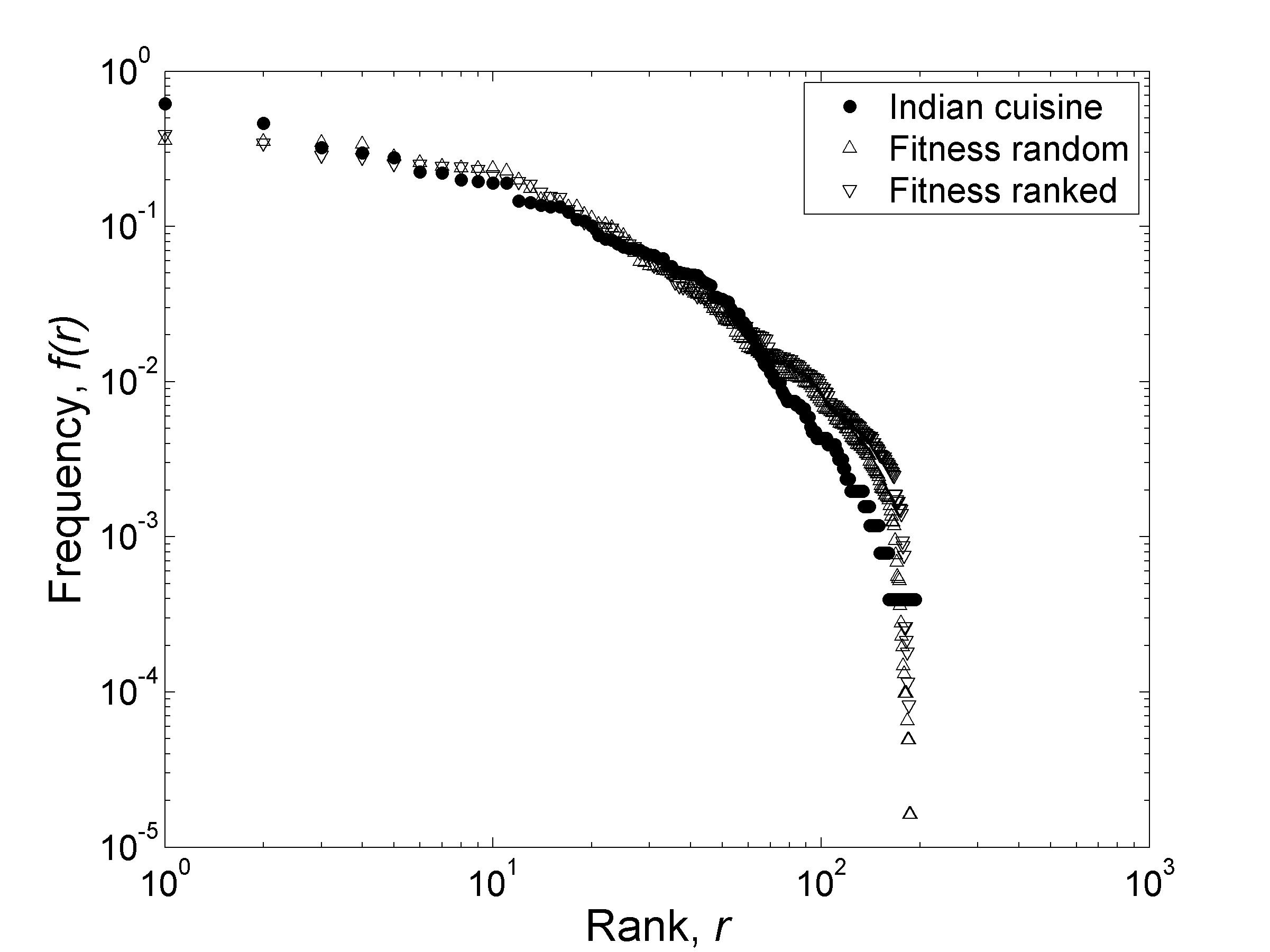}
\caption{\label{fig:freq_rank_ic}Frequency-rank distributions of ingredients for Indian cuisine and corresponding copy-mutate model (`Fitness random' having randomly assigned fitness values for ingredients) and its variant (`Fitness ranked') with frequency scaled fitness values for ingredients. The distribution in both cases match closely to that of the real world cuisine. The logged data was fitted with equation ($f(x) = a*exp^{bx}$) giving value of coefficient $b$ as 0.5664 for Indian cuisine, 0.5873 for fitness random model and 0.4809 for the fitness ranked model.}
\end{center}
\end{figure}

Fig.~\ref{fig:freq_rank_ic} shows the frequency-rank distributions for Indian cuisine and corresponding copy-mutate models of both random fitness values and empirical frequency based fitness values. The figure indicates that the frequency-rank distribution pattern gets reproduced by both the models. This can further be emphasized by looking at the coefficient values for the exponential fitting of the curves, as listed in figure caption.

\begin{figure*}
\begin{center}
\includegraphics[scale=.42]{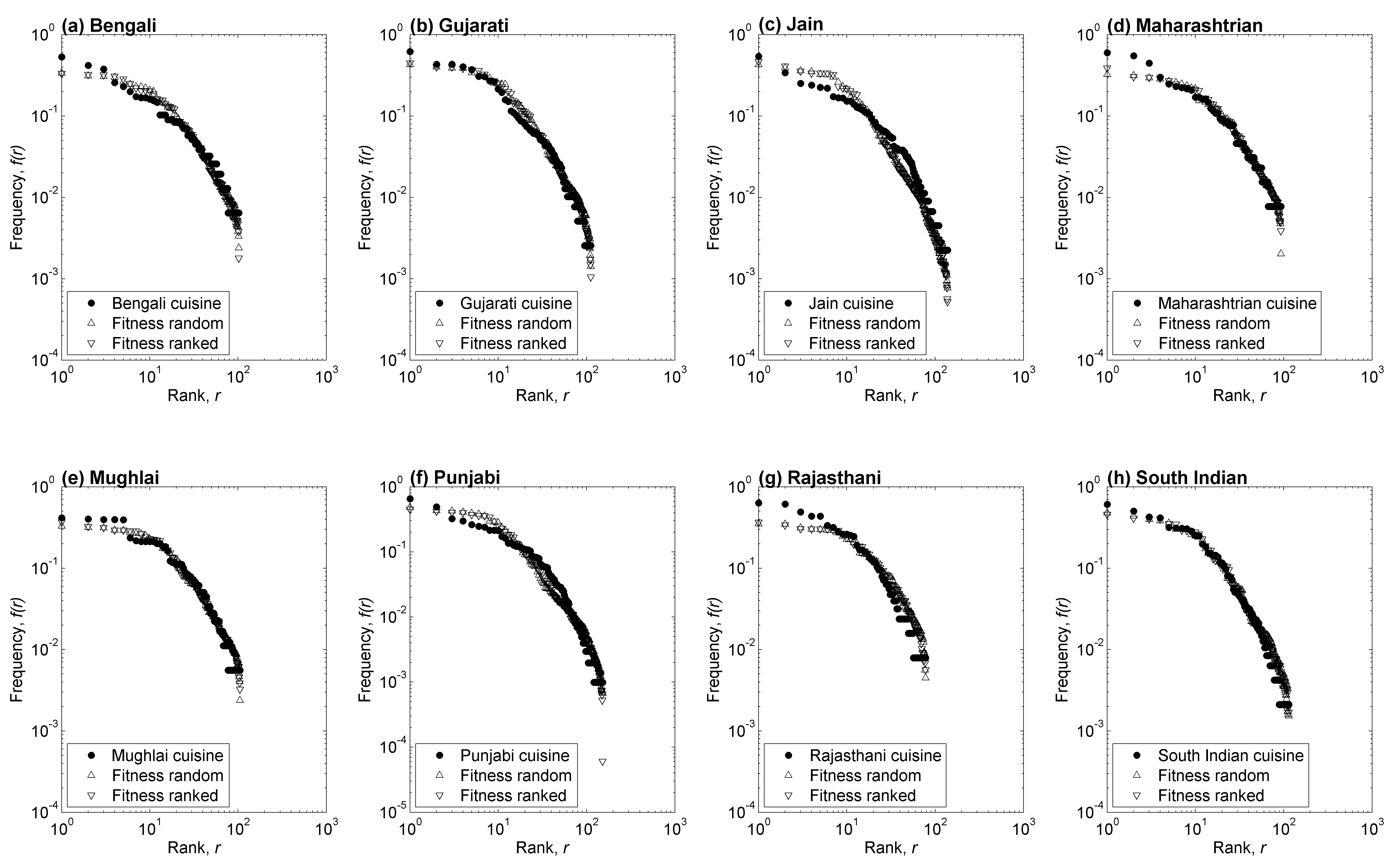}
\caption{\label{fig:freq_rank_all_rc}Frequency-rank distributions for all eight regional cuisines and corresponding models. The distributions for models closely match that of the empirical data in each one consistently.}
\end{center}
\end{figure*}

The reproduction of frequency-rank could also be seen generically across all the eight regional cuisines in India, as shown in Fig.~\ref{fig:freq_rank_all_rc}. All the curves were fitted with equation $f(x) = a*exp^{bx}$ and corresponding $b$ coefficient values are presented in table~\ref{tab:fit_coeff_b}.

\begin{table}[!]
\caption{\label{tab:fit_coeff_b}%
Values of fitting coefficient `$b$' for all eight regional Indian cuisines and corresponding models.}
\begin{ruledtabular}
\begin{tabular}{lcdr}
\textrm{Cuisine}&
\textrm{RC}\footnote{Regional cuisine}&
\textrm{CM-FRand}\footnote{Copy-mutate fitness random}&
\textrm{CM-FRank}\footnote{Copy-mutate fitness ranked}\\
\colrule
Bengali & 0.4302 & 0.4842 & 0.4608\\
Gujarati & 0.5204 & 0.507 & 0.5241 \\
Jain & 0.4784 & 0.4947 & 0.4758 \\
Maharashtrian & 0.463 & 0.4709 & 0.4601 \\
Mughlai & 0.5347 & 0.4794 & 0.4794 \\
Punjabi & 0.5702 & 0.4783 & 0.5075 \\
Rajasthani & 0.5761 & 0.505 & 0.5382 \\
South Indian & 0.5696 & 0.5321 & 0.4997 \\
\end{tabular}
\end{ruledtabular}
\end{table}

\subsection{Food pairing pattern}
The notion of food pairing is well-known in culinary science. The food pairing hypothesis, that two ingredients sharing common flavor compounds taste well together, has been widely researched upon in previous studies~\cite{Ahn2011,Jain2015}. Beginning from pairs of ingredients and corresponding number of shared flavor compounds ($N$), calculating the average flavor sharing of a recipe ($N_s^R$) and that of a cuisine (average $N_s$) has also been well-established in these studies. For our current study we have made use of these calculation methodologies only in order to test our model's capability of flavor pairing effect regeneration.

\begin{figure*}
\begin{center}
\includegraphics[scale=.4]{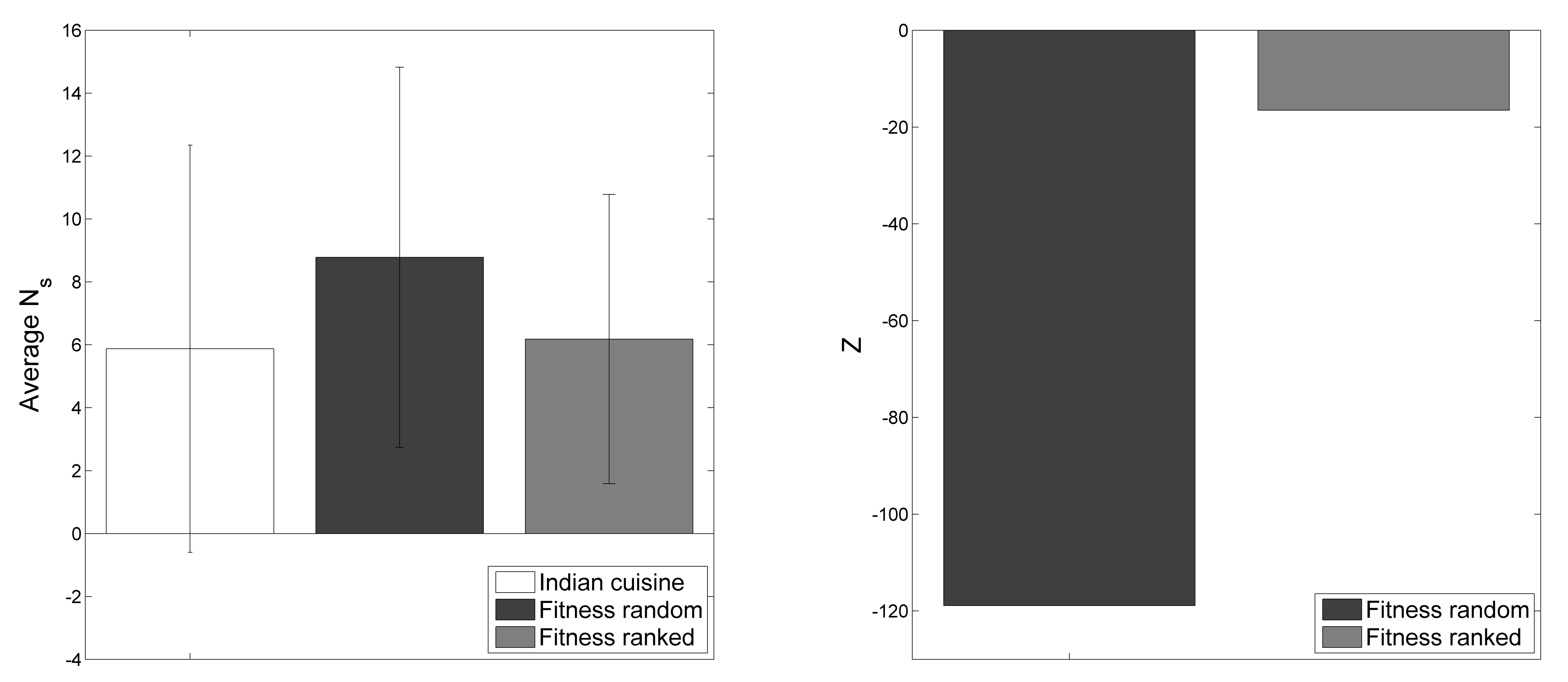}
\caption{\label{fig:avg_Ns_IC}Average $N_s$ values of Indian cuisine and the two copy-mutate models (`Fitness random' and `Fitness ranked') and their $z$-scores. The model with ranked values of fitness produces closer average $N_s$ value compared to that of the random one. Corresponding statistical significance is shown by the $z$-score.}
\end{center}
\end{figure*}

While studying and regenerating the frequency-rank distribution of ingredients itself is statistically interesting enough, can such a model reproduce empirically observed flavor sharing patterns as well? To answer this question, we began with comparing the average $N_s$ values~\cite{Ahn2011,Jain2015} of both the copy-mutate cuisines with that of the Indian cuisine (Fig.~\ref{fig:avg_Ns_IC}). As shown, the model cuisine with occurrence based fitness of ingredients has a closer average $N_s$ value to that of the Indian cuisine, while the random fitness based model cuisine's average $N_s$ is much higher indicating that certain fitness domain can produce a better model in terms of overall flavor effect observed. This further established that certain highly used ingredients play a vital role in defining the characteristic of the cuisine.

\begin{figure*}
\begin{center}
\includegraphics[scale=.4]{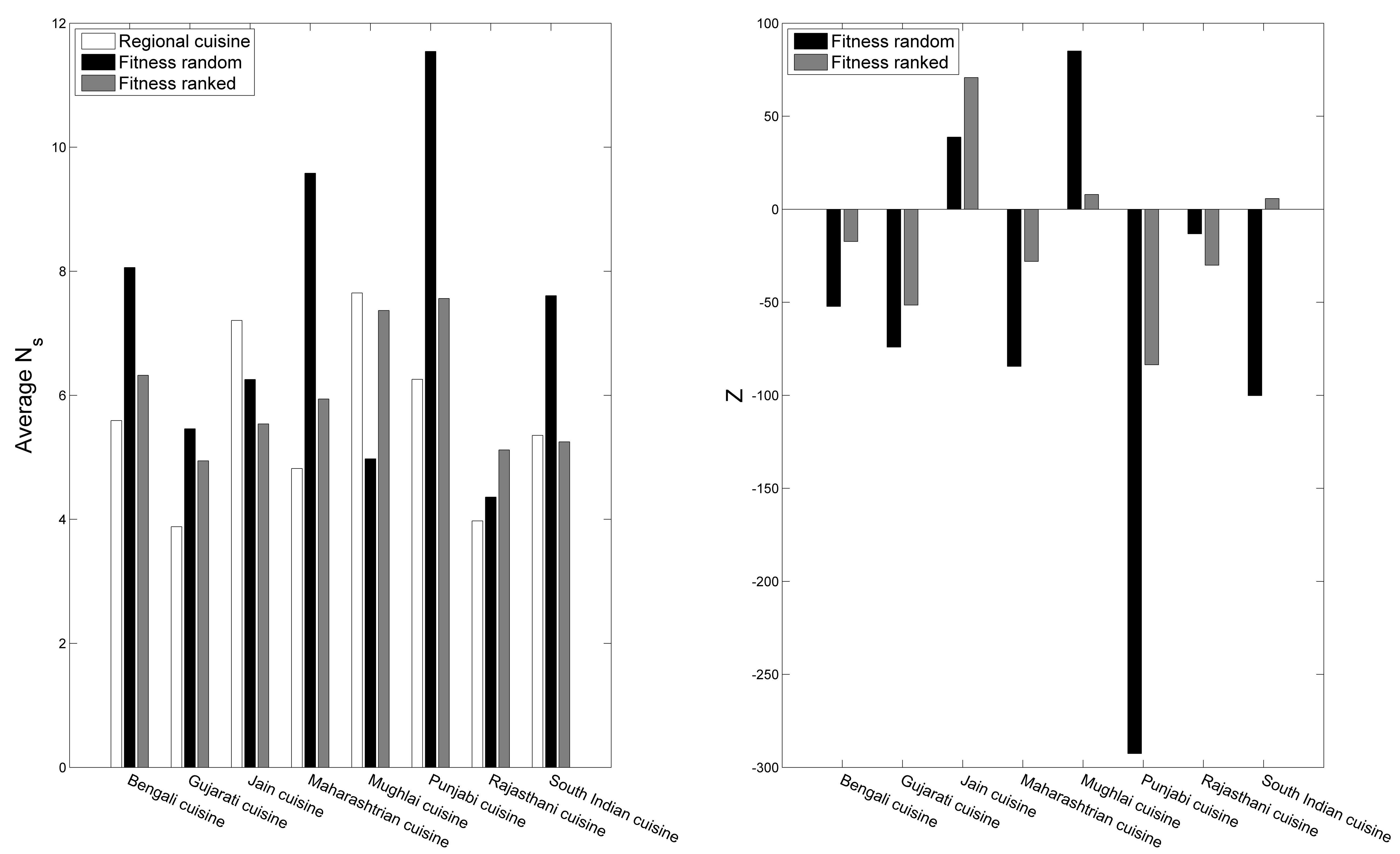}
\caption{\label{fig:avg_Ns_all_RC}Average $N_s$ values of eight Indian regional cuisines, their copy-mutate models (`Fitness random' and `Fitness ranked') and their $z$-scores.}
\end{center}
\end{figure*}

The model was applied for all the eight regional cuisines so as to check its applicability across cuisines. Interestingly, for all regional cuisines, barring \emph{Jain} and \emph{Rajasthani}, the copy-mutate model with ranked fitness values of ingredients produced better results for average $N_s$. This is shown in Fig.~\ref{fig:avg_Ns_all_RC}.

\begin{figure*}
\begin{center}
\includegraphics[scale=.42]{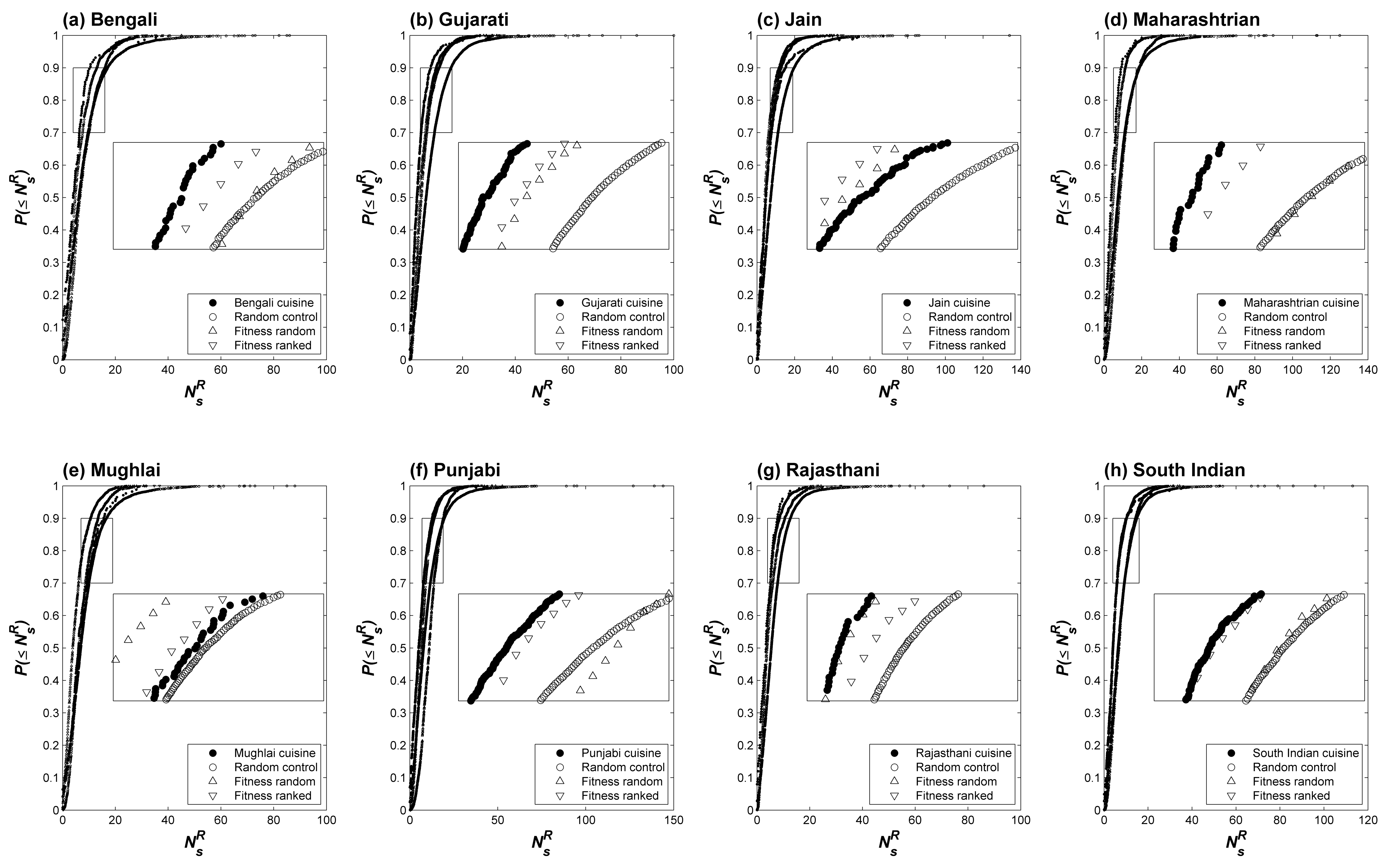}
\caption{\label{fig:PNs_vs_Ns_all_RC}Cumulative distribution of $P(N_s^R)$ vs $N_s^R$ values of eight regional cuisines of India and their copy-mutate models.}
\end{center}
\end{figure*}

However, the average $N_s$ over entire cuisine is not necessarily a strong reflector of the underlying flavor pattern. So we look a level deeper and check the recipe level distribution of the $N_s^R$ values. As indicated by Fig.~\ref{fig:PNs_vs_Ns_all_RC}, the distribution of $N_s^R$ values over the cuisine (average of 24 sets in case of copy-mutate model) also gets closer to empirical distribution as we move from random fitness to occurrence based fitness domain. The model with random fitness, as expected, shows the pattern closer to that of the uniform random model of the cuisine (one in which recipe-size distribution was preserved but recipes composed of uniformly selected ingredients). This further enhances our observation that the model with specific fitness domain is capable of producing a cuisine comparable with Indian cuisine in terms of flavor profile as well.

\begin{figure*}
\begin{center}
\includegraphics[scale=.5]{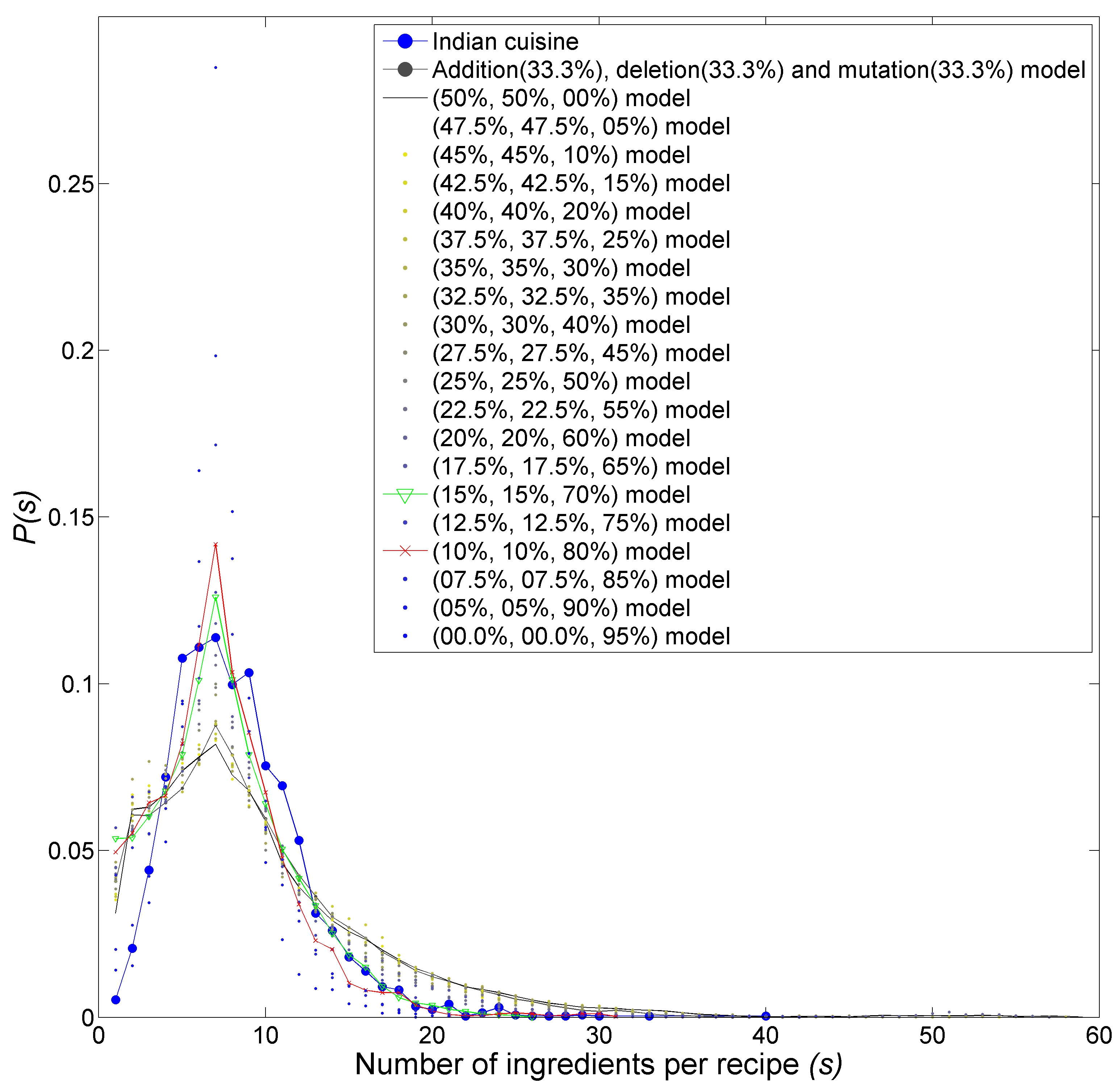}
\caption{\label{fig:rcp_size_dist} Recipe size distributions for the Indian cuisine and trials of CMAD model with varying probabilities of addition, deletion and mutation of recipes.}
\end{center}
\end{figure*}

As another alternative, we tried to create a model which could recreate the recipe-size distribution of a cuisine with inclusion of probabilistic addition and deletion of ingredients from recipes instead of only replacement of ingredients (mutation). Though exact replication of recipe-size distribution could not be achieved, as Fig.~\ref{fig:rcp_size_dist} indicates, certain probability values of addition, deletion and mutation get the recipe-size distribution similar to that of the real cuisine. 

Interestingly, mutation seemed to have been the dominating factor in evolution of the Indian cuisine as only those trials of the CMAD model gave better results for the recipe-size distribution which had higher probability value for mutation to occur compared to addition or deletion. If historical data were available, this observation could prove useful in generating the phylogenetic tree of recipes.

\section{Conclusions}
Quantitative as well as data-centric analysis of world cuisines has caught attention of physicists recently. We had earlier shown that the Indian cuisine is unique in its strong negative food pairing pattern~\cite{Jain2015}. In this study we focused on models~\cite{Kinouchi2008a} of eight Indian regional cuisines to probe for mechanisms that might have been dominant in their evolution over centuries. Our models highlight the role of `ingredient frequency' in rendering the characteristic food pairing pattern of Indian recipes. Further, we looked for the processes that are central to the recipes-size distribution and observed that phenomenon of mutation (change of one ingredient with another) very well explains the observed pattern. Our models and corresponding studies highlight the possibility of having an algorithmic way to suggest novel ingredient combinations as recipes~\cite{Varshney2013a} while still maintaining the flavor signature of a cuisine.

\begin{acknowledgments}
G.B. acknowledges the seed grant support from Indian Institute of Technology Jodhpur (IITJ/SEED/2014/0003). A.J. thanks the Ministry of Human Resource Development, Government of India as well as Indian Institute of Technology Jodhpur for scholarship.
\end{acknowledgments}



\begin{thebibliography}{14}%
\makeatletter
\providecommand \@ifxundefined [1]{%
 \@ifx{#1\undefined}
}%
\providecommand \@ifnum [1]{%
 \ifnum #1\expandafter \@firstoftwo
 \else \expandafter \@secondoftwo
 \fi
}%
\providecommand \@ifx [1]{%
 \ifx #1\expandafter \@firstoftwo
 \else \expandafter \@secondoftwo
 \fi
}%
\providecommand \natexlab [1]{#1}%
\providecommand \enquote  [1]{``#1''}%
\providecommand \bibnamefont  [1]{#1}%
\providecommand \bibfnamefont [1]{#1}%
\providecommand \citenamefont [1]{#1}%
\providecommand \href@noop [0]{\@secondoftwo}%
\providecommand \href [0]{\begingroup \@sanitize@url \@href}%
\providecommand \@href[1]{\@@startlink{#1}\@@href}%
\providecommand \@@href[1]{\endgroup#1\@@endlink}%
\providecommand \@sanitize@url [0]{\catcode `\\12\catcode `\$12\catcode
  `\&12\catcode `\#12\catcode `\^12\catcode `\_12\catcode `\%12\relax}%
\providecommand \@@startlink[1]{}%
\providecommand \@@endlink[0]{}%
\providecommand \url  [0]{\begingroup\@sanitize@url \@url }%
\providecommand \@url [1]{\endgroup\@href {#1}{\urlprefix }}%
\providecommand \urlprefix  [0]{URL }%
\providecommand \Eprint [0]{\href }%
\providecommand \doibase [0]{http://dx.doi.org/}%
\providecommand \selectlanguage [0]{\@gobble}%
\providecommand \bibinfo  [0]{\@secondoftwo}%
\providecommand \bibfield  [0]{\@secondoftwo}%
\providecommand \translation [1]{[#1]}%
\providecommand \BibitemOpen [0]{}%
\providecommand \bibitemStop [0]{}%
\providecommand \bibitemNoStop [0]{.\EOS\space}%
\providecommand \EOS [0]{\spacefactor3000\relax}%
\providecommand \BibitemShut  [1]{\csname bibitem#1\endcsname}%
\let\auto@bib@innerbib\@empty
\bibitem [{\citenamefont {Kinouchi}\ \emph {et~al.}(2008)\citenamefont
  {Kinouchi}, \citenamefont {Diez-Garcia}, \citenamefont {Holanda},
  \citenamefont {Zambianchi},\ and\ \citenamefont {Roque}}]{Kinouchi2008a}%
  \BibitemOpen
  \bibfield  {author} {\bibinfo {author} {\bibfnamefont {O.}~\bibnamefont
  {Kinouchi}}, \bibinfo {author} {\bibfnamefont {R.~W.}\ \bibnamefont
  {Diez-Garcia}}, \bibinfo {author} {\bibfnamefont {A.~J.}\ \bibnamefont
  {Holanda}}, \bibinfo {author} {\bibfnamefont {P.}~\bibnamefont {Zambianchi}},
  \ and\ \bibinfo {author} {\bibfnamefont {A.~C.}\ \bibnamefont {Roque}},\
  }\href {\doibase 10.1088/1367-2630/10/7/073020} {\bibfield  {journal}
  {\bibinfo  {journal} {New Journal of Physics}\ }\textbf {\bibinfo {volume}
  {10}},\ \bibinfo {pages} {1} (\bibinfo {year} {2008})},\ \Eprint
  {http://arxiv.org/abs/0802.4393} {arXiv:0802.4393} \BibitemShut {NoStop}%
\bibitem [{\citenamefont {Ahn}\ \emph {et~al.}(2011)\citenamefont {Ahn},
  \citenamefont {Ahnert}, \citenamefont {Bagrow},\ and\ \citenamefont
  {Barab\'{a}si}}]{Ahn2011}%
  \BibitemOpen
  \bibfield  {author} {\bibinfo {author} {\bibfnamefont {Y.-Y.}\ \bibnamefont
  {Ahn}}, \bibinfo {author} {\bibfnamefont {S.~E.}\ \bibnamefont {Ahnert}},
  \bibinfo {author} {\bibfnamefont {J.~P.}\ \bibnamefont {Bagrow}}, \ and\
  \bibinfo {author} {\bibfnamefont {A.-L.}\ \bibnamefont {Barab\'{a}si}},\
  }\href {\doibase 10.1038/srep00196} {\bibfield  {journal} {\bibinfo
  {journal} {Scientific Reports}\ }\textbf {\bibinfo {volume} {1}},\ \bibinfo
  {pages} {1} (\bibinfo {year} {2011})},\ \Eprint
  {http://arxiv.org/abs/arXiv:1111.6074v1} {arXiv:arXiv:1111.6074v1}
  \BibitemShut {NoStop}%
\bibitem [{\citenamefont {Jain}\ \emph {et~al.}(2015)\citenamefont {Jain},
  \citenamefont {Rakhi},\ and\ \citenamefont {Bagler}}]{Jain2015}%
  \BibitemOpen
  \bibfield  {author} {\bibinfo {author} {\bibfnamefont {A.}~\bibnamefont
  {Jain}}, \bibinfo {author} {\bibfnamefont {N.}~\bibnamefont {Rakhi}}, \ and\
  \bibinfo {author} {\bibfnamefont {G.}~\bibnamefont {Bagler}},\ }\href
  {arxiv.org/abs/1502.03815} {\bibfield  {journal} {\bibinfo  {journal}
  {arXiv:1502.03815}\ ,\ \bibinfo {pages} {1}} (\bibinfo {year} {2015})},\
  \Eprint {http://arxiv.org/abs/1502.03815} {arXiv:1502.03815} \BibitemShut
  {NoStop}%
\bibitem [{\citenamefont {Pollan}(2014)}]{Pollan2014}%
  \BibitemOpen
  \bibfield  {author} {\bibinfo {author} {\bibfnamefont {M.}~\bibnamefont
  {Pollan}},\ }\href@noop {} {\emph {\bibinfo {title} {{Cooked: A Natural
  History of Transformation}}}}\ (\bibinfo  {publisher} {Penguin Books},\
  \bibinfo {year} {2014})\ p.\ \bibinfo {pages} {480}\BibitemShut {NoStop}%
\bibitem [{\citenamefont {Appadurai}(2009)}]{Appadurai2009}%
  \BibitemOpen
  \bibfield  {author} {\bibinfo {author} {\bibfnamefont {A.}~\bibnamefont
  {Appadurai}},\ }\href {\doibase 10.1017/S0010417500015024} {\bibfield
  {journal} {\bibinfo  {journal} {Comparative Studies in Society and History}\
  }\textbf {\bibinfo {volume} {30}},\ \bibinfo {pages} {3} (\bibinfo {year}
  {2009})}\BibitemShut {NoStop}%
\bibitem [{\citenamefont {Sherman}\ and\ \citenamefont
  {Billing}(1999)}]{Sherman1999}%
  \BibitemOpen
  \bibfield  {author} {\bibinfo {author} {\bibfnamefont {P.~W.}\ \bibnamefont
  {Sherman}}\ and\ \bibinfo {author} {\bibfnamefont {J.}~\bibnamefont
  {Billing}},\ }\href@noop {} {\bibfield  {journal} {\bibinfo  {journal}
  {Bioscience}\ }\textbf {\bibinfo {volume} {49}},\ \bibinfo {pages} {453}
  (\bibinfo {year} {1999})}\BibitemShut {NoStop}%
\bibitem [{\citenamefont {Zhu}\ \emph {et~al.}(2013)\citenamefont {Zhu},
  \citenamefont {Huang}, \citenamefont {Zhang}, \citenamefont {Zhang},
  \citenamefont {Zhou},\ and\ \citenamefont {Ahn}}]{Zhu2013}%
  \BibitemOpen
  \bibfield  {author} {\bibinfo {author} {\bibfnamefont {Y.-X.}\ \bibnamefont
  {Zhu}}, \bibinfo {author} {\bibfnamefont {J.}~\bibnamefont {Huang}}, \bibinfo
  {author} {\bibfnamefont {Z.-K.}\ \bibnamefont {Zhang}}, \bibinfo {author}
  {\bibfnamefont {Q.-M.}\ \bibnamefont {Zhang}}, \bibinfo {author}
  {\bibfnamefont {T.}~\bibnamefont {Zhou}}, \ and\ \bibinfo {author}
  {\bibfnamefont {Y.-Y.}\ \bibnamefont {Ahn}},\ }\href {\doibase
  10.1371/journal.pone.0079161} {\bibfield  {journal} {\bibinfo  {journal}
  {PloS one}\ }\textbf {\bibinfo {volume} {8}},\ \bibinfo {pages} {e79161}
  (\bibinfo {year} {2013})}\BibitemShut {NoStop}%
\bibitem [{\citenamefont {Birch}(1999)}]{Birch1999}%
  \BibitemOpen
  \bibfield  {author} {\bibinfo {author} {\bibfnamefont {L.~L.}\ \bibnamefont
  {Birch}},\ }\href {\doibase 10.1146/annurev.nutr.19.1.41} {\bibfield
  {journal} {\bibinfo  {journal} {Annual Review of Nutrition}\ }\textbf
  {\bibinfo {volume} {19}},\ \bibinfo {pages} {41} (\bibinfo {year}
  {1999})}\BibitemShut {NoStop}%
\bibitem [{\citenamefont {Ventura}\ and\ \citenamefont
  {Worobey}(2013)}]{Ventura2013a}%
  \BibitemOpen
  \bibfield  {author} {\bibinfo {author} {\bibfnamefont {A.~K.}\ \bibnamefont
  {Ventura}}\ and\ \bibinfo {author} {\bibfnamefont {J.}~\bibnamefont
  {Worobey}},\ }\href {\doibase 10.1016/j.cub.2013.02.037} {\bibfield
  {journal} {\bibinfo  {journal} {Current Biology}\ }\textbf {\bibinfo {volume}
  {23}},\ \bibinfo {pages} {R401} (\bibinfo {year} {2013})}\BibitemShut
  {NoStop}%
\bibitem [{\citenamefont {Breslin}(2013)}]{Breslin2013}%
  \BibitemOpen
  \bibfield  {author} {\bibinfo {author} {\bibfnamefont {P.~A.~S.}\
  \bibnamefont {Breslin}},\ }\href {\doibase 10.1016/j.cub.2013.04.010}
  {\bibfield  {journal} {\bibinfo  {journal} {Current biology : CB}\ }\textbf
  {\bibinfo {volume} {23}},\ \bibinfo {pages} {409} (\bibinfo {year}
  {2013})}\BibitemShut {NoStop}%
\bibitem [{\citenamefont {Varshney}\ \emph
  {et~al.}(2013{\natexlab{a}})\citenamefont {Varshney}, \citenamefont
  {Varshney}, \citenamefont {Wang},\ and\ \citenamefont
  {Myers}}]{Varshney2013}%
  \BibitemOpen
  \bibfield  {author} {\bibinfo {author} {\bibfnamefont {K.~R.}\ \bibnamefont
  {Varshney}}, \bibinfo {author} {\bibfnamefont {L.~R.}\ \bibnamefont
  {Varshney}}, \bibinfo {author} {\bibfnamefont {J.}~\bibnamefont {Wang}}, \
  and\ \bibinfo {author} {\bibfnamefont {D.}~\bibnamefont {Myers}},\
  }\href@noop {} {\enquote {\bibinfo {title} {{Flavor Pairing in Medieval
  European Cuisine : A Study in Cooking with Dirty Data}},}\ } (\bibinfo {year}
  {2013}{\natexlab{a}}),\ \Eprint {http://arxiv.org/abs/arXiv:1307.7982v1}
  {arXiv:arXiv:1307.7982v1} \BibitemShut {NoStop}%
\bibitem [{\citenamefont {Dalal}(2014)}]{Dalal2014}%
  \BibitemOpen
  \bibfield  {author} {\bibinfo {author} {\bibfnamefont {T.}~\bibnamefont
  {Dalal}},\ }\href {http://www.tarladalal.com/} {\enquote {\bibinfo {title}
  {{Tarladalal.com}},}\ } (\bibinfo {year} {2014})\BibitemShut {NoStop}%
\bibitem [{\citenamefont {Burdock}(2010)}]{Burdock2010}%
  \BibitemOpen
  \bibfield  {author} {\bibinfo {author} {\bibfnamefont {G.~A.}\ \bibnamefont
  {Burdock}},\ }\href@noop {} {\emph {\bibinfo {title} {{Fenaroli's Handbook of
  Flavor Ingredients}}}},\ \bibinfo {edition} {6th}\ ed.\ (\bibinfo
  {publisher} {CRC Press},\ \bibinfo {year} {2010})\ p.\ \bibinfo {pages}
  {2159}\BibitemShut {NoStop}%
\bibitem [{\citenamefont {Varshney}\ \emph
  {et~al.}(2013{\natexlab{b}})\citenamefont {Varshney}, \citenamefont {Pinel},
  \citenamefont {Varshney}, \citenamefont {Bhattacharjya}, \citenamefont
  {Schoergendorfer},\ and\ \citenamefont {Chee}}]{Varshney2013a}%
  \BibitemOpen
  \bibfield  {author} {\bibinfo {author} {\bibfnamefont {L.~R.}\ \bibnamefont
  {Varshney}}, \bibinfo {author} {\bibfnamefont {F.}~\bibnamefont {Pinel}},
  \bibinfo {author} {\bibfnamefont {K.~R.}\ \bibnamefont {Varshney}}, \bibinfo
  {author} {\bibfnamefont {D.}~\bibnamefont {Bhattacharjya}}, \bibinfo {author}
  {\bibfnamefont {A.}~\bibnamefont {Schoergendorfer}}, \ and\ \bibinfo {author}
  {\bibfnamefont {Y.-M.}\ \bibnamefont {Chee}},\ }\href
  {http://arxiv.org/abs/1311.1213} {\ ,\ \bibinfo {pages} {1} (\bibinfo {year}
  {2013}{\natexlab{b}})},\ \Eprint {http://arxiv.org/abs/1311.1213}
  {arXiv:1311.1213} \BibitemShut {NoStop}%
\end{thebibliography}

%

\end{document}